\documentclass[11pt,dvips]{article} 

\begin{document}

\title{  { \Large { \bf 
Origin of Mass.  Mass and Mass-Energy Equation 
 from Classical-Mechanics  Solution
 } }}

\vspace{-0.4cm}
\author{
J. X. Zheng-Johansson$^{1}$ and 
P-I. Johansson$^{2}$ 
\\
{\small  {\it 1:   Inst. of Fundamental Physics Research, 611 93 Nyk\"oping, Sweden }}\\
 {\small {\it 2:  Dept. of Neutron Research, Uppsala University, 611 82 Nyk\"oping, Sweden}}
} 
\date{January, 2006}

\maketitle
\def\jm{\jmath}
\def\Scal{{\cal S}}
\def\velsub{{_{\mbox{\scriptsize${v}$}}         }}
\def\Thetam{\mit{\Theta}}
\def\r{{\mbox{\tiny${{\cal R}}$}}}
\def\re{{\mbox{\tiny${R}$}}}
\def\Fmed{F_{{\rm a.med}}}
\def\med{{\rm med}}
\def\Lw{L_{\varphi}}

\def\Mcal{{\cal{M}}}
\def\AR{\widetilde{A}}
\def\eng{\varepsilon}
\def\Eng{{\cal E}}
\def\Efb{{\bf E}}
\def\Bfb{{\bf B}}
\def\Bb{{\bf B}}
\def\Fb{{\bf F}}
\def\Rb{{\bf R}}
\def\Kb{{\bf K}}
\def\kb{{\bf k}}
\def\Nw{N_{b\phiv}}
\def\pdg{j}

\def\Ksub{{\mbox{\tiny${K}$}}}

\def\w{\omega{}}
\def\W{{\mit \Omega}}
\def\Lw{L_{\rm \Ac}}
\def\lam{\lambda}
\def\Lam{{\mit \Lambda}}
\def\Ac{ \varphi}
\def\phiv{\varphi}
\def\pmm{p}
\def\Pm{P}
\def\sang{{\cal Q}}
\def\th{\theta}
\def\pdgsup{{\mbox{\scriptsize${\pdg}$}}}
\def\dagsup{{\mbox{\tiny${\dagger}$}}}
\def\ddagsup{{\mbox{\tiny${\ddagger}$}}}
\def\aph{\alpha}
\def\D{\Delta}
\def\Taum{{\mit \Tau}}
\def\Nu{{\cal V}}
    \def\csub{{\mbox{\tiny${{\rm c}}$}}}
\def\hc{{h_\csub}}
\def\hbarc{{\hbar_\csub}}
\def\pd{\partial}
\def\lf{\left}
\def\rt{\right}
\def\NLamLw{N_{\Lam \varphi}}
\def\NlamLw{N_{\lam \varphi}}
\def\fsub{{_{^{f}}}}
\def\rhov{\rho_0}
\def\arm{{\rm a}}
\def\brm{{\rm b}}
\def\crm{{\rm c}}
\def\drm{{\rm d}}
\def\erm{{\rm e}}
\def\g{\gamma{}} 
\def\nud{\Nu_d{}}
\def\td{{\mit \Tau}_d{}}
\def\lamd{{\mit \Lambda}_d{}}
\def\Tau{\Gamma{}}

\begin{abstract} 
We establish the classical wave equation for a particle formed of a massless oscillatory elementary charge  generally also traveling, and the resulting electromagnetic waves, of a generally Doppler-effected angular frequency $\w$, in the vacuum in three dimensions. We obtain from the solutions the total energy of the particle wave to be  $\eng=\hbarc\w$,  $2\pi \hbarc$ being a function expressed in wave-medium parameters and identifiable as the Planck constant. In respect to the train of the waves as a whole traveling at the finite velocity of light $c$, $\eng=mc^2$ represents thereby the translational kinetic energy of the wavetrain,  $m=\hbarc\w/c^2$ being its  inertial mass and thereby the inertial mass  of the particle. Based on the solutions we also write down a set of semi-empirical equations for the particle's de Broglie wave parameters. From the standpoint of overall modern experimental indications we comment on the origin of mass implied by the solution. 
\end{abstract}

\section{Introduction} \label{Sec-intr}
Inertial mass ($m$) is a notion originally introduced into being through Isaac Newton's three laws of motion (1687) and is one elementary property of all matter substances. The inertial mass was postulated on phenomenological grounds by Albert Einstein in 1905 to be related with particle's total energy ($\eng$) as $\eng=mc^2$; the relationship has subsequently been broadly demonstrated in experiments. Nevertheless, the origin and nature of inertial mass remain up to the present one of the longest-standing unsolved problems  of physics. 

We have recently derived  based on overall experimental observations a particle formation scheme\cite{Ref1,RefunifB,Ref134,Ref245,Ref1-abs}, by which  a basic material particle like an electron, proton, positron and others, briefly, is formed of an oscillatory massless elementary charge with either sign, together with the electromagnetic waves generated by it in the vacuum. The vacuum is filled of entities termed as vacuuons, each consisting of a positive and negative elementary charge bound with a binding energy $\sim 10^7$ J, and 
will, in the presence of an applied field,
be electrically polarized and thereby induced with a shear elasticity. Such a substantial vacuum is, to remark especially here, consistently pointed to by overall experimental observations including the Michelson-Morley type of experiments, in particular  the modern experimental pair processes. In the  pair annihilation 
$e^-+ e^+ \rightarrow \gamma + \gamma $  
for instance, the two emitted gamma ($\gamma$) rays  carry the energy ($2 M_e c^2=2 \times 511 $ keV) converted from masses  of the electron $e^-$ and positron $e^+$  only,  whereas the Coulomb potential energy $V_c = -\frac{e^2}{4 \pi \epsilon_0 r_0}$ between the two particles of  charges $+e$ and $-e$, with a separation  distance $r_0$ just before annihilation, are not released according to the observations. Merely as a requirement of conservation of the energy, $V_c$ and certain carriers of it  which we call a $p$- and $n$- vaculeon, must inevitably remain in the vacuum after the annihilation. A substantial vacuum has practically been increasingly realized as realistic in some physics branches today that remain yet to employ a corresponding relativity theory.

A particle as given from classical-mechanics solutions (introduced in full in \cite{Ref1,RefunifB}, condensedly on two subtopics in \cite{Ref134,Ref245}) for the afore-mentioned model system  has the overall observational properties including the possession of a characteristic charge, mass and spin, being a wave, and obeying the de Broglie relations, Schr\"odinger equation\cite{Ref134}, Galilean-Lorentz transformation and Newton's law of gravitation \cite{Ref245} under corresponding conditions, among others.  In this paper we illustrate in a self-contained manner the solutions for such a particle for its  total energy, mass, the corresponding  mass-energy equation, and the equations, in semiempirical terms here, for Planck constant and de Broglie wave parameters. In the end we comment on the origin of mass, and also the analogy and contrast of our proposition with the contemporary proposal of the Higgs mechanism.

\section{Equation of motion} \label{Sec-2a}
Consider in the vacuum a  single massless oscillatory elementary charge $q$, $q=+e$ or $-e$, which being point-like relative to the  environment, 
having an oscillatory frequency $\W_q$, 
 and in general also traveling at a velocity $v$  in $+X$-direction. $v$ may be generally large such that $(v/c)^2>>0$ where $c$ being the velocity of light,
referred to as unclassic-velocity or relativistic regime. The charge is endowed with a free-particle kinetic energy $\Eng_q$ at its creation, thereafter responsible for the oscillation in the vacuum; $\Eng_q$  cannot be dissipated in the vacuum where there presents no lower energy levels for the charge to decay to except in a pair annihilation. The charge's radiation displacement, assuming relatively small here, can be well described by a sinusoidal function, $u_q^\jm =A_q^\jm  \sin (\W_q T)$, $A_q^\jm $ being the displacement amplitude, 
$j=\dagger $ and $\ddagger$ indicating the effect of $v$ as viewed from $+X$ and $-X$ directions. 
The orientation of $u_q^\jm  $ is of a random nature as the result of the influences of environmental fields which being random. Let $u_q^\jm   $ be, until equation (\ref{eq-Cc}), along the $Z$-axis in a given time interval for the discussions below.

The point-like charge will  owing to its oscillation generate electromagnetic waves, propagated at the velocity of light $c$ in radial directions in three dimensions here, in a  spherical enclosure of radius $R=L/2$ of hard-wall; a boundary is a  model representation of material substances that realistically always  present about any objects, here the charge and its immediate surrounding vacuum, near or farther in the physical world. We consider until Sec. \ref{Sec-3a} a duration which is  brief so the charge, assuming just  passing the origin $X=0$ along the $X$-axis, is essentially standing still, but is  long so as to enable the waves to travel many loops across the diameter $L$,  a situation typically realized for $v<<c$; the source motion effect is, until the explicit solutions in Sec. \ref{Sec-3a}, formally indicated by suffixes $\dagger $ and $ \ddagger$ after the wave variables. A radial  path $\Rb$ is in spherical polar coordinates described by $\Rb(R, \th,\phi)$ making an angle $\th$ at the $Z$-axis and an azimuthal angle $\phi$ at the $X$-axis. A radial  wave propagated in $+R$ or $-R$ direction, labeled by $\jmath=\dagger$ or $\ddagger$,  in the Maxwellian field picture, to reflect briefly, consists of a transverse radiation electric field $E^\jm(R,\th,\phi,T)= \frac{\mu_0 q \W_q^2 A_q^\jm   \sin \th}{ 4 \pi R}    \sin[k^\jm{} R \mp\w^\jm T] $,   and  magnetic field $B^\jm(R,\th,\phi,T)=E^\jm(R,\th,\phi,T)/c$ (see e.g. \cite{Ref1} or \cite{Grifitth} for a detailed derivation), where $k^\jm$ being the normal mode wavevector               and $\w^\jm$ angular frequency.

It is a familiar long established fact that, to the source charge here, its alternating electric field manifests hand in hand with its {\it mechanical displacement}. In an  analogous fashion, 
to as far as their apparent phenomenological correspondence goes only in this paper (for a systematic justification see \cite{Ref1}), 
we can represent  the radiation $E^\jm$ field  
in the vacuum medium  
with a corresponding mechanical transverse ($\theta$-direction) displacement $u^\jm(R,\th,\phi,T)$ of it,  
as a response to the source disturbances;  
the vacuum is simultaneously polarized by the static field of the source charge and induced with an elasticity  (cf. earlier comments). 
Here $\jm=\dagger  $ and $ \ddagger$, indicating as earlier two partial radial waves described by $u^\dagsup$ and $u^\ddagsup$ are  generated and initially propagated in $+R$ and $-R$ directions.
More explicitly, we consider a cone region,  or cone chain,  of the medium along the chosen wave path $R$, of a differential  solid angle 
$$\refstepcounter{equation} \label{eq-sang1}
d \sang= \sin \th d \th d \phi   
                         \eqno(\ref{eq-sang1})
 $$ 
As  a heuristic means only in this paper, we represent the cone chain 
 as consisting of coupled oscillators each made of a thin shell  of thickness $b$ between two spherical surfaces  at $R $ and $ R+b$, of a volume $b R^2 d \sang$, and thus mass 
$$\refstepcounter{equation} \label{eq-Mas1}
d \Mcal_b (R,\th,\phi)= \rho_0 b R^2 d \sang = \rho_0 b R^2  \sin \th d \th d \phi 
            \eqno(\ref{eq-Mas1})
$$ 
Where  $\rho_0$ is the volume (dynamic) mass density of the vacuum. The inter-oscillator distance is accordingly $b$. (As an information only here, the $b$ corresponding to the linear size of the building units of the vacuum termed  vacuuons, is 
shown in  \cite{Ref1} to be of the order of a Planck length).   

 Along the cone chain we look at  a segment $\D L$ being  small such that  its cross sectional area $S(R,\theta,\phi)$, and thus linear mass density $\rho_l(R,\theta,\phi)=S \rho_0$, is to a good approximation constant, which is exact in the limit $b/\D L \rightarrow 0$. $\D R$ is yet sufficiently larger than $b$ so as to maintain the medium therein a continuum. The transverse ($\theta$-direction) displacements of the segment  $u^{\dagsup}$ and $u^{\ddagsup}$ as produced and initially propagated in $+R$- and $-R$-directions, are accordingly uniform across $\D R$ in the aforesaid manner. The elastic vacuum along the cone chain, upon the deformation described by $u^{\dagsup}$ and $u^{\ddagsup}$, is subject to a tensile force,  $F_\r(\Rb,T)$; $F_\r=c^2 \rho_l$  as the solution will verify below. 
 $F_\r$ is across the small segment uniform accordingly. 
Yet $\D L$ is tilted from $R$ differently,  at angles  $\Thetam^\jm $ and $\Thetam^\jm+\D \Thetam^\jm $  at its ends $R$  and $ R+\D R$; either angle is in turn dependent on the wave variables and thus (see below) the wave travel directions, hence suffixed by $\jm=\dagger,$  $\ddagger$.  The geometry determines $\sin \Thetam^\jm = \frac{\D u^\jm}{\D L}$,  and $\sin (\Thetam^\jm+\D \Thetam^\jm )-\sin (\Thetam^\jm) \simeq \Thetam^\jm = \frac{\partial^2 u^\jm}{\partial R^2} \D R$ for  $u^\jm$ small; the transverse ($\theta$-) component of the net force on $\D L$  is then to a good approximation:
$$\displaylines{
\refstepcounter{equation}\label{eq-Frp}
\hfill
\D F_{\r.t}^{ j}
= F_\r [\sin (\Thetam^\jm+\D \Thetam^\jm )-\sin (\Thetam^\jm)]
=\rho_l c^2 \frac{\partial^2 u^\jm }{\partial R^2} \D R 
\hfill (\ref{eq-Frp}) 
}  $$
Applying Newton's second law for the segment $\D L$, $ \simeq \D R$, of mass $\rho_l \D R$ gives its equation of motion  in the transverse $\theta$-direction: 
$\D {F^\jm}_{\r.t} = \frac{ \partial ^2 \varphi^\jm}{\partial T^2} \rho_l \D R $.
 With (\ref{eq-Frp}) for its left-hand side,  dividing then $\rho_l \D R$ through, we obtain the final equation of motion for unit length-density  of the chain, or equivalently the classical wave equation for the particle's  component total waves:
   $$\displaylines{\refstepcounter{equation} \label{eq-eqmabs-a}
\label{eq-CME1}
\hfill \qquad\qquad   
c^2 \nabla^2 u^\jm (R,\th,\phi,T) -\frac{\pd^2 u^\jm (R,\th,\phi,T)}{\pd T^2}  =0,   \qquad \jm=\dagger, \ddagger 
\hfill (\ref{eq-eqmabs-a})
}$$

\section{Solutions}
\subsection{Local wave displacement and energy}
Within each small segment (\ref{eq-eqmabs-a}) has a plane wave solution, $u^\jm(R,\th,\phi,T) =\widetilde{A}{}^\jm   
 (R,\th,\phi,T)$ $\sin[k^\jm{} R \mp\w^\jm T ] $. Here  $\widetilde{A}{}^\jm  $ is the displacement  amplitude;  $k^\jm$ is the normal mode wavevector  and $\w^\jm$ angular frequency; $k^\jm R=\kb^\jm\cdot \Rb $; $\w^\jm=k^\jm c$, where $c=\W_q b$, as a standard result of  wave mechanics for  the continuum limit. $\widetilde{A}{}^\jm  $ is, as with the $\Mcal_b$ above,  dependent on   $R, \theta$ when going from segment to segment. The energy flux is however  along the  cone chain  constant; utilizing this character and sampling at two locations ($b$ and $R$) on the chain, with  some further simple algebras,  we get $\widetilde{A}{}^\jm   =\frac{b\sin \th A_q^\jm  }{R}  $, and accordingly 
$$\displaylines{\refstepcounter{equation} \label{eq-Ap1} \label{eq-ux1} 
\hfill  
u^\jm(R,\th,\phi,T)=\frac{b\sin \th A_q^\jm  }{R}  
\sin[k^\jm{} R \mp\w^\jm T  ]     \qquad \jm=\dagger, \ddagger 
                            \hfill (\ref{eq-ux1})
}$$
$u^\jm(R,\th,\phi,T)$,  similarly as the $E^\jm(R,\th,\phi,T)$ earlier, is maximum at $\th =\pi/2$ and  zero at $\th=0$,  and is independent of $\phi$; the field has a donut ring stereo contour about the symmetry $Z$-axis. 

Each oscillator of mass $d \Mcal_b$ and displacement $u^\jm$, given by (\ref{eq-Mas1}) and (\ref{eq-ux1}),  has a mechanical energy: 
$$\displaylines{
d \eng_b^\jm(R,\th,\phi,T) = 
\frac{1}{2} d \Mcal_b (R,\th,\phi) [\dot{u^\jm}^2(R,\th,\phi,T) + 
{\w^\jm}^2  {u^\jm}^2(R,\th,\phi,T)    ]    
\cr
\hfill \qquad\qquad 
=\frac{1}{2} d \Mcal_b (R,\th,\phi) {\w^\jm }^2 \lf(\frac{b\sin \th A_q^\jm  }{R} \rt)^2 \qquad\qquad\qquad  \hfill (\arm) 
}$$
Integrating (\arm) over the total solid angle $4\pi$ we obtain  the total wave energy passing a spherical shell of thickness $b$ at radius $R$ 
$$ \refstepcounter{equation}\label{eq-Eb}
\eng^\jm_b= \int_{\th=0}^\pi  \int_{\phi =0}^{2\pi}
d \Eng^\jm(R,\th,\phi,T) = \frac{1}{2}\Mcal_ b {\w^\jm}^2 {A}^2, \qquad \jm=\dagger, \ddagger  \eqno(\ref{eq-Eb})
$$
Where $\Mcal_ b=\frac{4}{3} \pi b^3 \rho_0 $, 
$A=A_q^\jm /(2\pi k^\jm b)$;
         $A$ can be readily verified to be independent of $v$.  

\subsection{Apparent linear chain representation. Effective plane waves }
(\ref{eq-Eb}) shows that, $ \eng^\jm_b$ is  independent of $R$. So {\it apparently}, the constant energy $\eng^\jm_b$ is effectively as if conveyed by a total {\it plane wave} propagated in a {\it linear} chain along the symmetry  $X$-axis, composed of coupled identical oscillators each of a mass $\Mcal_b$  and displacement amplitude $A$. In terms of  equivalence in linear momenta, this (effective) total plane wave consists in turn of two opposite traveling component plane waves in $+X$- and $-X$- directions;  their displacements as functions of $(X,T)$ may be effectively represented by those of the partial waves 
of  (\ref{eq-ux1}) along the $X$-axis, i.e.  
$\pm \sin(k^\jm X \mp\w^\jm T)$ in the two directions.
Accordingly  the spherical enclosure   reduces to a one-dimensional box of side $L$ centered at $X=0$ along the $X$-axis. In all, the radial waves are altogether equivalently in the aforesaid fashion described by  the two plane waves propagated in a one-dimensional box along the $X$-axis in $+X$- and $-X$-directions:
$$\refstepcounter{equation} \label{eq-uX1}
\left\{  {u^{\dagsup}(X,T)        \atop u^{\ddagsup}(X,T)}  \right\}= \pm A  \sin(k^\jm X \mp\w^\jm T), \qquad \jm=\dagger, \ddagger      
                \eqno(\ref{eq-uX1}) 
$$
Their full stretches on the wave path give at any instant two wavetrains, traveling at the speed of light $c$ in opposite directions. The two wavetrains present always in pair and describe together the dynamics of the total system, being thus said conjugate with each other. 

\subsection{Wavetrain length. Total wave energy}
Suppose each wavetrain winds about $L$ in $j^\jm$ loops, giving a total wavetrain length $\ell_\phiv^\jm=j^\jm L $,  $\jm=\dagger,\ddagger$ as before. The number of oscillators  contained in $\ell_\phiv^\jm$ 
is  
$$\refstepcounter{equation}\label{eq-Nw}
\Nw^\jm = \frac{\ell_\phiv^\jm}{b}
                   \eqno(\ref{eq-Nw})
$$
$\eng_b^\jm$ multiplied with $\Nw$, and in turn with a normalization factor $C_o$, gives then the normalized  total wave energy   
$$\displaylines{
\hfill \eng^\jm 
= C_o \Nw^\jm   \eng^\jm_b
\hfill (\brm)
}$$
Here, $C_0$ takes into account the effect that the oscillation explores all directions in $4\pi$ solid angle over a sufficiently long time, hence 
$$\refstepcounter{equation}\label{eq-Cc} 
C_o=1/4 \pi         \eqno(\ref{eq-Cc}) 
$$
Substituting  with (\ref{eq-Eb}), (\ref{eq-Nw}) and (\ref{eq-Cc}),  (\brm) explicitly writes
$$\displaylines{
\hfill \eng^\jm 
=\frac{\ell_\phiv^\jm}{8 \pi b} 
\Mcal_b {\w^\jm}^2 A^2 \hfill (\ref{eq-Er2}) 
\cr
\mbox{Or alternatively } \hfill 
\cr   \refstepcounter{equation} \label{eq-Er2} 
\hfill 
\eng^\jm = \hbarc \w^\jm                   \hfill (\ref{eq-Er2}\arm) 
\cr 
{\rm where \ } \hbarc
= \frac{\ell^\jm_\phiv}{8 \pi}  \frac{\Mcal_b
}{b} (c \frac{2\pi}{\lam^\jm}) A^2
= \frac{1}{4}\frac{ \NlamLw^\jm \lam^\jm
}{\lam^\jm} \Scal \rho_0 c A^2. \
{\rm Or} \hfill
\cr
\refstepcounter{equation} \label{eq-hp}
\hfill 
\hbarc= \frac{1}{4}  \NlamLw \Scal \rhov c A^2.
\hfill (\ref{eq-hp})
}$$
In the above,  $\lam^\jm$ ($=2\pi /k^\jm$) is the wavelength,  and $\NlamLw= \ell^\jm_\phiv/\lam^\jm$ gives the number of wavelengths in $\ell^\jm_\phiv$. $\lam^\jm$ and $\ell^\jm_\phiv$ ($\jm=\dagger, \ddagger$) are in general each dependent on $v$  but their ratio $\NlamLw$ will maintain a constant in the case where  $v$ may be large so that $(v/c)^2>>0$, but is  yet small such that it does not alter the energy level of the total wave, i.e. to a $\NlamLw'$ different from $\NlamLw$ (this latter energy can be estimated for typical particle systems like an orbiting atomic electron to be much greater than the  $v$ associated particle kinetic energy). $\Scal=\frac{\Mcal_b}{b \rho_0}=\frac{4}{3} \pi b^2$ is  the effective cross sectional area of the wave path.  

(\ref{eq-Er2}a) is seen to be formally just Max Planck's energy equation  for electromagnetic wave (1900);  $2\pi \hbarc=\hc$ can be accordingly identified as the Planck constant $2\pi \hbar=h $. That is
$$\displaylines{
\refstepcounter{equation} \label{eq-hph}
\hfill \hc= h;\hfill \hbarc=\hbar \hfill (\ref{eq-hph})
}$$
Hence (\ref{eq-hp}) is an expression for $h$  in terms of wave-medium parameters; although the above is not a proof of the constancy of $h$.

In the special  case that the source is oscillatory about a fixed location, i.e.  $v=0$, we evidently have two identical opposite-traveling conjugate wavetrains. We denote the corresponding dynamic variables  with the corresponding lowercase letters, a denotation method adopted throughout this paper. E.g. 
$\Eng^\jm|_{v=0} =\eng^\jm$, $\Pm^\jm|_{v=0} =\pmm^\jm$, $L_\phiv^\jm{}|_{v=0} =\ell_\phiv^\jm$, $\lam^\jm|_{v=0} =\Lam$, $\tau^\jm|_{v=0} =\Taum$, $\w^\jm|_{v=0} =\W$, $k^\jm|_{v=0} =K$. 
The respective  pair of extensive quantities, being identical ($ \Eng^{\dagsup}= \Eng^{\ddagsup}$, $ \Pm^{\dagsup}= \Pm^{\ddagsup}$ and $\Lw^{\dagsup}= \Lw^{\ddagsup}$) 
and each equal to one half of the total, can  simply add up, giving the total quantities: 
$\frac{1}{2} \Eng^{\dagsup}+\frac{1}{2}\Eng^{\ddagsup} = \Eng$, $\frac{1}{2} \Pm^{\dagsup}+\frac{1}{2}\Pm^{\ddagsup} = \Pm$, and    
$\frac{1}{2} \Lw^{\dagsup}+\frac{1}{2}\Lw^{\ddagsup} = \Lw$.   

\subsection{Doppler-effected wave variables owing to source motion}\label{Sec-3a}
We now go beyond the brief moment in which the source charge was essentially standing still (i.e. oscillating about a fixed location, at the origin $R=0$ in the illustration), over to a macroscopic time scale, at  which  the effect of the translational motion of  the source, with a velocity $v$ in the $+X$-direction as earlier, needs be explicitly considered.
 The velocity of light $c$ (as measured by an observer at rest in the vacuum, 
let be fixed in frame $S$, in which  we shall remain throughout this paper), similar as any ordinary mechanical waves,  is not altered by the source motion. On the other hand, by a pure mechanical virtue the moving source surpasses and recedes from the waves it generates to its right and left each by a distance $\Taum v$,  where $\Taum$ ($=\Lam/c$) is the wave period and $\Lam$ wavelength due to the source when being localized,  $v=0$, all referring to measurement made in $S$.  Consequently the wavelengths of the moving source  are  shortened and elongated from $\Lam$ to
$$ \displaylines{  
\refstepcounter{equation}\label{eq-Lam1} 
 \hfill 
\lam^{\dagsup} =  \lf(1-{v/c}\rt) \Lam  \hfill (\ref{eq-Lam1}\arm) \hfill  
\lam^{\ddagsup} = \lf(1+{v/c}\rt) \Lam 
                            \hfill  (\ref{eq-Lam1}\brm)
}$$
Accordingly the angular frequencies $\w^\jm=c/\lam^\jm$ are displaced to: 
$$ \displaylines{  
\refstepcounter{equation}\label{eq-W1}
 \hfill  \w^{\dagsup}= \frac{\W}{1-v/c}   \hfill (\ref{eq-W1}\arm)  \hfill
 \w^{\dagsup}= \frac{\W}{1+v/c} 
                                 \hfill(\ref{eq-W1}\brm)
}$$
With (\ref{eq-W1}) in (\ref{eq-Er2}\arm) we have the corresponding Doppler-effected total wave energies 
$$ \displaylines{ \refstepcounter{equation}  \label{eq-Er2A} 
\hfill 
\eng^{\dagsup}  ={\hbarc} \w^{\dagsup}     \hfill (\ref{eq-Er2A}\arm)  
\hfill 
\eng^{\ddagsup}  ={\hbarc}  \w^{\ddagsup}                                
                          \hfill (\ref{eq-Er2A}\brm)
}$$
 ${\hbarc} $ being  as given in (\ref{eq-hp}).

\section{Wavetrain dynamics}
\label{Sec-4x}
For the afore-represented waves we now examine the basic dynamics and relations of the  {\it wavetrains as a whole}. For this we first recall a familiar result of Maxwellian electrodynamics  stating that the energy  $\eng^\jm$ and linear momentum $\pmm^\jm$ of an electromagnetic wave $j$ ($=\dagger, \ddagger$)  are related as 
$$\displaylines{\refstepcounter{equation}\label{eq-Pm}
\hfill  \pmm^{\dagsup}=\frac{\eng^{\dagsup}}{c} \quad (\arm)
 \hfill   
\pmm^{\ddagsup}=\frac{\eng^{\ddagsup}}{c} \quad (\brm)
\hfill (\ref{eq-Pm}) 
}$$
Where $\pmm^\jm = \Delta T  F^\jm_{m}$ corresponds to the impulse of the Lorentz magnetic force $F^\jm_m$ exerted by  wavetrain $j$ on a test charge  accumulated during time $\Delta T$. That the linear momentum $\pmm^\jm$ of an electromagnetic wave takes the form (\ref{eq-Pm}), instead of $p^\jm= \frac{2\eng^\jm}{c}$ as would be for a material particle moving at a velocity denoted by the same symbol $c$ here, is well understood through the electrodynamics solution itself to be as a result that, to a charge they pass, only the $E$ field does work and not the $B$ field.  

If we now {\it overlook the details } of the rapid wave oscillation, and focus on the energy and momenta conveyed by them, we then perceive the two wavetrains just as {\it two rigid objects}, traveling at the speed of light $c$, in $+X$ and $-X$ directions. Each wavetrain, having a velocity of travel $c$ which is {\it finite} as opposed to infinite, and having a translational kinetic energy $\eng^\jm$  and linear momentum $\pmm^\jm$ as given above, is apparently an {\it inertial} system. Each wavetrain must thereby have a corresponding inertial mass, $m^\jm$, where $\jm=\dagger, \ddagger$. It thus follows that $m^\jm$ is   according to {\it Newton's second law} related to $\pmm^\jm$ as:
 $$\displaylines{\refstepcounter{equation} 
\label{eq-P1ab} 
\hfill       \pmm^{\dagsup}=m^{\dagsup} c 
\hfill  (\ref{eq-P1ab}\arm) \qquad\qquad 
  \pmm^{\ddagsup}=m^{\ddagsup} c       
\hfill  (\ref{eq-P1ab}\brm)   
}$$
Eliminating $p^\jm$ between (\ref{eq-Pm}) and (\ref{eq-P1ab}) gives further 
 $$\displaylines{
\refstepcounter{equation} \label{eq-EM} 
\hfill       
                 \eng^{\dagsup}=m^{\dagsup} c^2 
\hfill     (\ref{eq-EM}\arm)\qquad\qquad  
                  \eng^{\ddagsup}=m^{\ddagsup} c^2    
\hfill  (\ref{eq-EM}\brm)   
}$$
Either wavetrain is traveling at the speed $c$ in many, $j^\jm$,  loops between the   boundaries being here fixed. By the dynamic virtue however, the center of mass of each wavetrain  is traveling at the velocity of the charge, $v$, as is reflected through the Doppler displacements in the wave variables. 

\section{Particle.  Particle total dynamics}\label{Sec-3}

We shall hereafter refer to  the  oscillatory charge together  with the wavetrains generated by it as a whole as a {\it particle}. The additional (here  translational) motion of the oscillatory charge represents thereby the (translational) motion of the  particle, with the velocity $v$. The wavetrains and the charge  represent accordingly now the internal components and their motions the internal processes of the particle. It is worth remarking that the mechanical energy of the wavetrain and that of the charge are two manifestations of the same energy, only one should be counted for at one time.

We will below be concerned with the dynamic variables of the particle.
The  conjugate waves are mutually stochastic, as follows from  their relative  phases being random, as a combined consequence of the instantaneous position of the source relative to the boundaries. 
By their this stochastic virtue and their  being sampled at one time,  the total dynamic variables in wave terms follow to be appropriately given by the geometric means of the respective variables of the conjugate wavetrains. 
For the extensive quantities like $\eng$, $\pmm$ and $m$, the total mean is in turn  twice the mean of per wavetrain. In sum, the wavelength,  angular frequency, the total energy, total momentum and the mass of the  (total wave of the) particle, traveling at velocity $v$ (here in $+X$-direction),  are given by 
$$\displaylines{  
\refstepcounter{equation}  \label{eq-EXlam}
\hfill \lam = \sqrt{\lam^{\dagsup}{}\lam^{\ddagsup}{}} \hfill (\ref{eq-EXlam})
\cr
\refstepcounter{equation}  \label{eq-EX4}
\hfill 
\w = \sqrt{\w^{\dagsup}{}\w^{\ddagsup}{}} 
                           \hfill (\ref{eq-EX4})
\cr
\refstepcounter{equation}  \label{eq-EX1}
\refstepcounter{equation}  \label{eq-EX2}
\refstepcounter{equation}  \label{eq-EX3}
\hfill  
\eng 
=2 \times  \sqrt{(1/2)\eng^{\dagsup} \cdot (1/2)\eng^{\ddagsup}}
= \sqrt{\eng^{\dagsup}\eng^{\ddagsup}} 
                             \hfill  (\ref{eq-EX1})
\cr
\hfill 
\pmm 
=2 \times  \sqrt{(1/2)\pmm^{\dagsup}{} \cdot (1/2)\pmm^{\ddagsup}{}} 
=  \sqrt{\pmm^{\dagsup}{} \pmm^{\ddagsup}{}} 
                             \hfill  (\ref{eq-EX2})
\cr
\hfill 
m 
= 2 \times \sqrt{(1/2)m^{\dagsup}{}\cdot (1/2) m^{\ddagsup}{}} 
= \sqrt{m^{\dagsup}{}m^{\ddagsup}{}} 
                            \hfill   (\ref{eq-EX3})
}$$  
Making the algebraic operations: (\ref{eq-Lam1}\arm)$\times $(\ref{eq-Lam1}\brm), (\ref{eq-W1}\arm)$\times $(\ref{eq-W1}\brm), (\ref{eq-Er2A}a) $\times$ (\ref{eq-Er2A}b), 
(\ref{eq-Pm}a) $\times$ (\ref{eq-Pm}b), (\ref{eq-P1ab}a) $\times$ (\ref{eq-P1ab}b) and (\ref{eq-EM}a) $\times$(\ref{eq-EM}b), sorting, and taking square roots through for each resultant equation, we have
$$\displaylines{ 
\hfill \sqrt{\lam^{\dagsup} \lam^{\ddagsup}} 
= \Lam\sqrt{1-(v/c)^2}; 
\hfill
\sqrt{\w^{\dagsup} \w^{\ddagsup}} = \frac{\W}{\sqrt{1-(v/c)^2}}; 
\hfill 
 \sqrt{\eng^{\dagsup}\eng^{\ddagsup}}  =\hbarc 
\sqrt{\w^{\dagsup}   \w^{\ddagsup}}; 
                                        \hfill 
\cr
\hfill \sqrt{\eng^{\dagsup}\eng^{\ddagsup}}=\sqrt{\pmm^{\dagsup}   \pmm^{\ddagsup}   }\ c; 
\hfill 
\sqrt{\pmm^{\dagsup} \pmm^{\ddagsup}} = \sqrt{m^{\dagsup} m^{\ddagsup}} \ c;      \hfill  
\sqrt{\eng^{\dagsup} \eng^{\ddagsup}} = \sqrt{m^{\dagsup} m^{\ddagsup}} \ c^2 
\hfill
}$$
Substituting into these with   
  (\ref{eq-EXlam})-(\ref{eq-EX3})
for the respective square root quantities we finally have:
$$\displaylines{
\refstepcounter{equation}\label{eq-lam2}
\hfill \qquad\quad
\lam = \frac{\Lam }{\g}   \qquad  \hfill(\ref{eq-lam2})
     \qquad\qquad {\rm where} \quad             
                \Lam=\sqrt{\Lam^\dagsup\Lam^\ddagsup}
                   \hfill (\ref{eq-lam2}\arm)
\cr 
\refstepcounter{equation}\label{eq-wW}
 \hfill
\w =\frac{2\pi c}{\lam}=\g \W   
                          \hfill(\ref{eq-wW})
\qquad\qquad {\rm where} \quad             
                          \W =\frac{2\pi c}{\Lam}
                                                 \  \hfill (\ref{eq-wW}\arm)
\cr
 \refstepcounter{equation}\label{eq-M1a} 
  \hfill  \eng = \hbarc \w =\g \Eng   \hfill (\ref{eq-M1a})
\qquad\qquad {\rm where} \quad 
                          \Eng=\hbar \W  \ \hfill(\ref{eq-M1a}\arm)
\cr
 \refstepcounter{equation} \label{eq-M1c} 
       \label{eq-E1}    
  \hfill \eng= \pmm c =\g \Eng \hfill (\ref{eq-M1c})
\qquad\qquad {\rm where} \quad
                     \Eng=\Pm c   \ \hfill (\ref{eq-M1c}\arm)
\cr 
\refstepcounter{equation}\label{eq-M1b} 
  \hfill
 \pmm =m c =\g P     \hfill (\ref{eq-M1b})
\qquad\qquad {\rm where} \quad
                          P =Mc    \hfill (\ref{eq-M1b}\arm)
\cr
\refstepcounter{equation}\label{eq-emc2}
\hfill 
   \eng  =m c^2=\g \Eng  \hfill (\ref{eq-emc2})
\qquad\qquad {\rm where} \quad
                       \Eng= Mc^2        \hfill (\ref{eq-emc2}\arm)                      
}$$
 Canceling $\eng$ between (\ref{eq-emc2}) and (\ref{eq-M1a}),  reorganizing, we obtain $m$, in turn with $m$ in (\ref{eq-M1b}) we obtain  $\pmm$, and from (\ref{eq-wW}) we obtain $\tau$,   
all in terms of the wave-medium parameters:
$$\displaylines{\refstepcounter{equation} \label{eq-M10}
\hfill 
m = \frac{\hbarc \w}{c^2}  = \g M 
                           \hfill(\ref{eq-M10}) 
\qquad\qquad  {\rm where} \quad
M= \frac{\hbarc \W}{c^2}        
\hfill  (\ref{eq-M10}\arm) 
\cr\refstepcounter{equation} \label{eq-P10}
\hfill \pmm =\hbarc k =\g \Pm  
                               \hfill (\ref{eq-P10}) 
\qquad\qquad {\rm where} \quad \Pm= \hbar K    
                        \hfill(\ref{eq-P10}\arm)
\cr
\refstepcounter{equation} \label{eq-kK1}
\hfill k=\frac{\w}{c}=  \g K   \hfill (\ref{eq-kK1}) 
\qquad\qquad {\rm where} \quad
                 K=\frac{ \W}{c}        
\hfill (\ref{eq-kK1}\arm) 
\cr\refstepcounter{equation} \label{eq-tT1}
\hfill 
\tau=\frac{2\pi}{\w} = \frac{\Taum}{\g} \hfill(\ref{eq-tT1}) 
\qquad\qquad {\rm where} \quad
                    \Taum=\frac{2\pi}{ \W}
\hfill (\ref{eq-tT1}\arm)
}$$
In the above, 
$\Lam$, $\W$, $\Eng$, $P$, $M$,  $K$ and $\Taum$ are the corresponding variables for the particle at rest in vacuum; and 
$$\displaylines{ 
\refstepcounter{equation} \label{eq-g}
\hfill \g = \frac{1}{\sqrt{1-(v/c)^2}} \hfill (\ref{eq-g})
}$$
(\ref{eq-M1a}) is seen to correspond to Max Planck's energy equation for the total particle wave here. (\ref{eq-emc2}) expresses just the mass-energy equivalence relation postulated by Albert  Einstein (1905), and is given in the above as a direct classical-mechanics  solution for the particle total wave.

All of the foregoing solution equations expressed in lowercase-letter variables are obtained for the particle in its motion ($X$-) direction. The corresponding equations in the perpendicular ($Y$-, $Z$-) directions  to the particle's motion can be written down similarly but with the  component velocities being zero, $v_x=v_y=0$. We thus find the total dynamic variables of the moving particle are in the perpendicular  directions unaffected by its  motion,  
remaining as expressed by the uppercase letters.

If $v$ is small so as to be within the non-relativistic or classic-velocity regime, $(v/c)^2 \rightarrow 0$, then  $\lim_{(v/c)^2 \rightarrow 0} \g = 1$. Accordingly the particle total dynamic variables  take now the limit values:  
$
\lim_{(v/c)^2 \rightarrow 0} \w  =\W$, \ 
$ \lim_{(v/c)^2 \rightarrow 0} m 
 $ $ =\lim_{(v/c)^2 \rightarrow 0}\frac{\g \hbarc  \W}{c^2} 
 $ $  =M$, \ 
   $\lim_{(v/c)^2 \rightarrow 0}\pmm = \hbarc   \lim_{(v/c)^2 \rightarrow 0} \g K  =\Pm$, \ 
$\lim_{(v/c)^2 \rightarrow 0}\eng = \hbarc   \lim_{(v/c)^2 \rightarrow 0} $ $ \g \W  =  \Eng 
$, and similarly for other variables. These show that at the classic-velocity limit $(v/c)^2\rightarrow 0$, the particle motion makes no contribution to the total particle quantities  $\w, m, \pmm$ and $\eng$ etc. which thereby identify with the respective total rest quantities, $\W, M, \Pm$ and $\Eng$, etc. This is seen to be a result that these total quantities are dependant on the particle's velocity only at a second order in $v^2$, i.e. $\propto (v/c)^2$.

\section{Particle dynamics. De Broglie variables}
\label{Sec-4}

Expanding $\g$ in Taylor series,   
(\ref{eq-emc2}) then writes:
$$\displaylines{
\eng
 = Mc^2 \lf[1+\frac{1}{2} \lf(\frac{v}{c}\rt)^2 + \frac{1\cdot 3}{2 \cdot 4} \lf(\frac{v}{c}\rt)^4    +\ldots     \rt] 
= Mc^2 + \eng_v 
\cr
{\rm where } \hfill 
\cr\refstepcounter{equation} \label{eq-Y2a}
\hfill 
\eng_v=\Eng_v \lf[         1+ \frac{3}{4} \lf(\frac{v}{c}\rt)^2 +\ldots \rt]\hfill(\ref{eq-Y2a})
\cr
\hfill \Eng_v=\lim_{(v/c)^2 \rightarrow}\eng_v = \frac{1}{2}Mv^2 \hfill (\ref{eq-Y2a}\arm)
}$$
Here $\eng_v$, being the difference between the total energies $\eng$ of the particle in motion 
and  $\Eng$ at rest, clearly  represents just the exact translational kinetic energy of the particle. $\Eng_v $ is the classic-velocity limit of $\eng_v$. Accordingly the exact linear momentum of the particle and its classic-velocity limit, respectively, are  
$\pmm_\velsub =mv$  
(this is related to the component-wave variables as \cite{Ref1}   $ \pmm_\velsub=  \sqrt{i mv (- imv)} $) and  $\Pm_\velsub = \lim_{(v/c)^2 \rightarrow 0}\pmm_\velsub = Mv$. 
Substituting into the last equation  and  (\ref{eq-Y2a}\arm) 
 respectively  with 
(\ref{eq-M10}) for $M$, we get at the classic-velocity limit
$$\displaylines{ \refstepcounter{equation} \label{eq-p5}
\hfill 
\Pm_\velsub=  \hbarc  \lf(\frac{v}{c}\rt)  \frac{2\pi}{\Lam}  \hfill (\ref{eq-p5})  
\cr
\refstepcounter{equation} \label{eq-p5b}
 \hfill    \Eng_v = \frac{1}{2} \hbarc  \lf(\frac{v}{c}\rt)^2    \W
                  \hfill (\ref{eq-p5b})
}$$

Now according to Louis  de Broglie's hypothesis (1923) the particle is associated with a de Broglie wavelength, $\Lam_d^{{\rm em}}$, wavevector $K_d^{{\rm em}} =2\pi/\Lam_d^{{\rm em}}$ and angular frequency $\W_d^{\rm em} =K_d^{\rm em} v$; $\Lam_d^{{\rm em}}$ and $\W_d^{{\rm em}}  $ are related with  
 $ \Pm_\velsub$ and $\Eng_v$ by the de Broglie relations: 
$$\displaylines{
\refstepcounter{equation} \label{eq-emPEfp} 
\refstepcounter{equation} \label{eq-emPEfpb} 
\hfill \Pm_\velsub  = \hbar \frac{ 2 \pi  }{ \Lam_d^{{\rm em}} }   \hfill (\ref{eq-emPEfp})  
\cr
\hfill
   \Eng_v =  \frac{1}{ 2} \hbar \W_d^{{\rm em}}  \hfill (\ref{eq-emPEfpb})
}$$
In the above the superscript "em" indicates the equations  (\ref{eq-emPEfp})--(\ref{eq-emPEfpb}) follow from the de Broglie hypothesis on an empirical basis (to distinguish from ones deduced in full in   \cite{Ref1}, briefly in \cite{Ref134}, from direct wavefunction solutions).  With $\hbarc = \hbar$ as given after (\ref{eq-hph}), canceling $\Pm_\velsub/\hbar$ between the empirical equation (\ref{eq-emPEfp}), and the classical-mechanics solution equation (\ref{eq-p5}), and  similarly $\Eng_v/\hbar$ between (\ref{eq-p5b}) and (\ref{eq-emPEfpb}), we obtain a set of  semiempirical expressions for the de Broglie wavelength and frequency,  and accordingly  wavevector and period:
$$\displaylines{\refstepcounter{equation} \label{eq-lmd} 
\hfill \qquad\qquad
  \Lam_d^{{\rm em}} = \lf(\frac{c}{ v}\rt)\Lam 
                      \hfill  ({\rm a}) \qquad\qquad
\cr
\hfill \qquad\qquad
  \W_d^{{\rm em}}
                = \frac{2 \pi v}{ \Lam_d^{{\rm em}}}
                =  \lf(\frac{v}{ c}\rt)^2  \W 
\hfill ({\rm b})\qquad\qquad
\cr  
\hfill \qquad\qquad
 K_d^{{\rm em}}
=\frac{2\pi}{ \Lam_d^{{\rm em}} } 
= \lf(\frac{v}{ c}\rt)K  
                          \hfill ({\rm c})        \qquad\qquad
\cr
\hfill \qquad\qquad
    \Taum_d^{{\rm em}} = \frac{1}{ \Nu_d^{{\rm em}}}
=  \lf(\frac{c}{ v}\rt)^2 \Taum
                   \hfill ({\rm d})  
\qquad  (\ref{eq-lmd})
}$$
For the last expressions of (\ref{eq-lmd}b)-(d) we used 
(\ref{eq-wW}\arm), (\ref{eq-kK1}\arm) and (\ref{eq-tT1}\arm). These expressions turn out to be completely identical to those L. de Broglie obtained based on a hypothetical phase wave\cite{deBroglie:1924} if that  wave is supposed to transmit its energy at the velocity of light $c$. 

\section{Discussion} 
\label{Sec-5}

(a) {\it The physics of mass-energy relation}: The foregoing classical-mechanics solution has yielded directly that a particle's mass $m$ is related to its total energy $\eng$ as $\eng =mc^2$, in the particular form of Newtonian translational kinetic energy of a {\it mass} $m$ traveling at the {\it velocity of light} $c$; and this  is not by accident. This is the direct consequence that the {\it mass} $m$ executing this motion is nothing else but a (total) electromagnetic wavetrain, and $c$ is its velocity of travel, governed by Newton's laws of motion.

(b) {\it The origin of mass}: We have shown in the above that the total energy $\eng$ of a particle equals the total mechanical (or electromagnetic)  energy of the total component waves constituting the particle. This energy can be apprehended as being required for the particle to counterbalance a certain {\it frictional force} opposed to the particle's total motion;  this underlies the origin of mass.  That such a frictional force is indeed exerted by the vacuum, whence the vacuum frictional force, will follow immediately once the vacuum is given a physical representation as a substantial medium and the waves as mechanical processes in it. For the total particle dynamic solutions we have actually heuristically employed such a vacuum in this paper  (for a systematic deductive representation of the vacuum see  \cite{Ref1}). The physics of this friction in the vacuum  will have no principle difference  from that in a material medium, the latter being in general well understood. 

The frictional force of the vacuum, a medium known to us to fill everywhere in space, must be universally exerted on  any objects moving in it---in terms of total motion. It has  evidently the same universality as the inertial mass. In Isaac Newton's laws of motion this vacuum frictional force is  alternatively but equivalently assigned as an intrinsic property of the object, the inertial mass, or the  inertial force when expressed in the unit of force.  

(c) The origin of mass suggested in (b) is in conformity  with the contemporary Higgs mechanism\cite{Higgs1964} (see e.g. one of recent reviews in  \cite{Froggatt2003}), in the sense that mass results from some massless charge in a viscous vacuum (this proposal), or alternatively from certain component moving  in a certain field (Higgs mechanism). However apparently, the two proposals for the origin of mass are associated with two distinct schemes for particle formation. The schemes differ, primarily perhaps, in that the present scheme works in real space, whilst Higgs' in momentum space as  the quantum field theories do. 

As shown in this paper a particle simply formed by a single oscillatory charge is an extensive entity---wave---by formation; and this formation is governed by the established laws of classical mechanics and not by imposition. The particle's de Broglie wave properties,  and its relativistic mass and (as shown in  \cite{Ref1}) length contraction, are the natural results of  Doppler effect with a solid experimental grounding. Also as shown in our other reports\cite{Ref1,RefunifB,Ref134,Ref245} 
 the classical-mechanics description of such particle  leads directly to the Schr\"odinger equation, a universal gravitational interaction, and other overall observed properties of basic material particles. Such  desirable features tend to have been more readily endorsed by the real-space representation of the present scheme.

(d) Our discussion here is in respect of the center-of-mass motion of an elementary charge with either sign, which forms accordingly the observed isolatable basic material particles like the electron and proton. But apparently, the same scheme is viable also for the formation of a material  "particle" from any fraction of charge, like the net charge of a composite particle, or at today's finest scale the $(2/3)e$ or $-(1/3)e$ of a quark. The only requirement is that the total motion of all charges involved is employed for  computing  the particle's total energy and mass;
the particle's center-of-mass motion is responsible for particle dynamics.

\end{document}